\documentstyle[aps,prl,multicol,epsf]{revtex}
\input epsf

\newcommand {\Bias}{f}
\newcommand {\DimensionlessBias}{{f^*}}
\newcommand {\DimensionlessVelocity}{{v^*}}
\newcommand {\CombinedEnergy}{E}
\newcommand {\SpringConstant}{k}

\begin{document}
\draft

\title{Thermal effects on atomic friction 
}
\author{Yi Sang, Martin Dub\a'e, and Martin Grant
}
\address{
Centre for the Physics of Materials, 
Physics Department, Rutherford Building,
McGill University, 3600 rue University, 
Montr\a'eal, Qu\a'ebec, 
H3A 2T8 Canada 
}

\maketitle

\begin{abstract}
We model friction acting on the tip of an atomic force microscope as it is
dragged across a surface at non-zero temperatures. We find that 
stick-slip motion occurs and that 
the average frictional force follows $|\ln v|^{2/3}$, where $v$
is the tip velocity.  This compares well to recent experimental work,
permitting the quantitative extraction of all microscopic parameters.
We calculate the scaled form of the average frictional force's
dependence on both temperature and tip speed as well as the form of the
friction-force distribution function.

\end{abstract}

\pacs{46.55.+d, 81.40.Pq, 64.60.My, 07.79.-v}

\begin{multicols}{2}

The study of friction between two surfaces presents numerous theoretical
and experimental challenges \cite{Bhushan_99,Persson_98}. Two
macroscopic surfaces interact through many asperities.  These asperities
are on an atomic length scale, so macroscopic friction is inextricably
linked with microscopic properties.  
A friction force microscope, the tip of which is in essence a single
asperity \cite{Mate_87,Bouhacina_97,Gnecco_00}, 
provides an ideal way to probe atomic friction by removing
the complication of multiple asperities. 
Under a constant
load, the low-velocity motion of the tip on a surface exhibits stick-slip
behavior as the moving tip hops over an atomically-defined 
substrate \cite{Tomlinson_29,Basel,Helman_94,Hamburg,Colchero_99}. 
In principle, such results can
give a quantitative description of friction at the microscopic level.
However this requires a theoretical understanding of the behaviour of
the tip on the substrate.  This is the purpose of this 
Letter.

Recent studies have argued that, due to thermal fluctuations, 
the lateral friction force $F$ has a
logarithmic dependence on velocity \cite{Bouhacina_97,Gnecco_00} $F
\propto {\rm const} + \ln v$ where $v$ is the velocity of the support.
This corresponds to linear creep between 
two 
surfaces in contact 
when the force acting on a substrate produces
a small constant potential bias \cite{Heslot_94}. 
But this is quite different from what occurs when 
a tip is dragged across the substrate: 
the potential bias is
continuously ramped up as the support is moved.  We will call this ramped
creep and will show that, at constant temperature $T$, 
\cite{GUNTON,KLEIN,Kurki_72,Jackel_74,Garg_95,FISHER}
\begin{equation} 
F \propto {\rm const} - T^{2/3} \, |\ln v/T|^{2/3}.
\label{2/3} 
\end{equation}  

It is straightforward to discriminate between linear and ramped creep.
Consider Figs.\ \ref{fig1}a and \ref{fig1}b 
where the combined potential of the atomic substrate
and the elastic tip is shown.  Locally, it is sufficient to consider
two wells. In that case, a useful analogy to (mean-field) nucleation at a
first-order phase transition occurs \cite{GUNTON,KLEIN}.  When the bias is
infinitesimally small, 
the energy, up to an additive constant, 
is well described locally by $\CombinedEnergy =
Fx +x^2 - x^4$, where $F$ is friction and $x$
is displacement.  Hence the barrier height is $\Delta \CombinedEnergy
\sim ({\rm const} - F)$, as for mean-field nucleation near the
coexistence curve.  In the steady state, the rate of thermal 
fluctuations is 
proportional to the velocity. Since the rate is itself proportional to 
the probability of a fluctuation over this barrier,  
one obtains linear creep:
the force depends logarithmically on the velocity.  This is however an
unlikely regime because the barrier is so high (see Fig.\ \ref{fig1}a).
In contrast, as the tip is dragged over the substrate and the bias
is ramped up, 
barrier-hopping fluctuations 
occur preferentially when the tip is close to
slipping at the top of the barrier. At this point, 
the local combined potential
can be written $\CombinedEnergy = F x - x^3$, where the origin of the $x$
axis has been conveniently shifted, and the second well has been replaced
by an absorbing boundary condition  (see Fig.\ \ref{fig1}b). 
Then we have $\Delta
\CombinedEnergy \sim ({\rm const} - F)^{3/2}$ and  Eq.\ (1) for 
ramped creep is obtained. This is similar to mean-field
nucleation near the spinodal \cite{GUNTON,KLEIN}, 
and is calculated, as well as observed experimentally, for, e.g., 
magnetic-flux fluctuations in superconducting quantum interference devices
\cite{Kurki_72,Jackel_74,Garg_95}, and in other contexts \cite{FISHER}.  
We give a detailed exposition of the algebra below.  First, however,
we solve a model numerically and show that the experimental regime
corresponds to our physical picture.

The system composed of the tip of the force microscope interacting
with a surface can be described by a modified Tomlinson model
\cite{Basel,Helman_94,Hamburg,Colchero_99}.  We thus consider a tip
with longitudinal and transverse coordinate positions $x$ and $z$ and
effective mass $M$. The tip moves under the influence of the 
surface adiabatic
potential $\CombinedEnergy(x,z)$, coupled 
elastically to the support of the microscope,
of coordinate $R (t)$ and $Z (t)$. For constant load, the distance $Z
(t) - z$ is fixed, 
so the
transverse coordinate plays no role, and $R
= vt$ with $v$ the constant speed at which the tip is dragged.
Including the random effect of 
substrate fluctuations, we have the
Langevin equation 
\begin{equation}
 M \frac{ d^2 x}{d t^2} + M \gamma  \frac { d x}{d t}
 + \frac { \partial \CombinedEnergy(R,x)}{\partial x} = \xi(t), 
\label{langevin}
\end{equation}
where $M$ is the mass of the tip, $\gamma$ is the 
microscopic friction coefficient, 
$\CombinedEnergy$ is the combined surface-tip potential,
and $\xi$ is a random noise satisfying the fluctuation-dissipation
relation $\langle \xi (t) \xi (t') \rangle = 2  M  \gamma 
k_B T \delta ( t -t')$, where the angular brackets denote an
average, and $k_B$ is Boltzmann's constant. 
The combined surface-tip potential has the form 
\begin{equation}
\CombinedEnergy(R,x) = 
\frac{\SpringConstant}{2} \, ( R (t) - x)^2 - U \cos
\left( \frac{2 \pi x}{a} \right)
\label{st_potential}
\end{equation}
where $U$ is the surface barrier potential height, $a$ is the lattice
constant, and $\SpringConstant$ is the elastic constant of the tip-support
coupling.  
There are a a series of potential
wells $\tilde{x} (R)$ given by the solution to
$\partial \CombinedEnergy (R, x) /\partial x |_{\tilde x} = 0$, 
with the instantaneous lateral friction force on the support $F(t)
= \SpringConstant \, (R(t)-x(t))$.  
A typical sequence of
the stick-slip motion of $F$ is shown in Fig.\ 1c.

The relevant time scales of the problem are $\gamma^{-1}$, the
viscous time scale for dissipation of energy from the tip
to the surface \cite{GAMMA}, $p_0 = \sqrt{M a^2/U}$, the oscillation
period of the mass in the surface potential, and the tip resonance 
period  $p_k= \sqrt{ M/\SpringConstant}$. 
These parameters are effective in the sense that their precise
value may depend on the load applied to the 
friction force 
microscope \cite{NEW_REFERENCE}.
The friction coefficient must be 
larger than the critical value 
needed to observe
overdamped stick-slip
behaviour at zero temperature. Within the Tomlinson model, this value
is $\sim 2 \omega_k$ \cite{Basel,Hamburg,Colchero_99}.
To first order in the deviations of 
the tip 
from an integer position,
the instantaneous force 
can be 
approximated as
$F (t)  \simeq [{\SpringConstant}/(1+\Omega_k^2)] R (t)
\equiv \tilde{\SpringConstant} R$, 
where $\Omega_k = p_0 / (2\pi p_k)$.

Equation (\ref{langevin})
was simulated using Ermak's algorithm \cite{Allen_90}
with parameters consistent with experimental conditions. The 
lattice parameter $a = 0.4  \, \mbox{nm}$ and the effective spring
constant $\tilde{\SpringConstant} = 0.86 \, \mbox{N/m}$ \cite{note2} 
were taken from the data of Ref.\ \cite{Gnecco_00}, 
obtained
under constant load $F_n = 0.65 \, \mbox{nN}$. 
The parameters
$U = 0.27 \, {\mbox eV}$ and $\gamma = 8.9  \times 10^5
\, \mbox{sec}^{-1}$ and spring resonance frequency
$(2 \pi p_k)^{-1} = 52 \mbox{MHz}$ were adjusted to obtain the best 
fit with the experimental data. 
The mass  $M = 8.7 \times 10^{-12} \, \mbox{kg}$ then follows. 
The tip oscillation frequency 
$2\, \pi\, p_{0}^{-1} = 1.1 \times 10^6 \, 
\mbox{sec}^{-1}$ is a comparatively small attempt frequency compared to
usual values in adatoms' surface diffusion \cite{diff_gen}, which is 
why thermally activated events with very small barrier height are
important. 

The numerical simulations were performed over a large range of 
velocities $ 5 \, \mbox{nm/sec} \leq v \leq 256 \, \mbox{$\mu$m/sec}$,
with 
several temperatures 
$ 53 \, \mbox{K} \leq T \leq 373 \, \mbox{K}$ and 
time steps $ 0.001  \, p_0  \leq \delta  t \leq  0.01  \, p_0$. 
 From the stick-slip motion of $F$, we extracted the 
average lateral
force as a function of velocity, for different temperatures.  This
is shown in Fig.\ \ref{fig1}d.  To test the prediction for ramped creep,
we first extracted the constant force in Eq.\ (\ref{2/3}), $F_c$, 
from the data at different temperatures, but fixed ratio of
the scaled velocity 
$v/T$, as shown in the top inset of Fig.\ 2.
With this constant removed, the prediction from Eq.\ (1) 
corresponds to a universal form, which is independent of temperature.  
Fig.\ 2 shows a very good scaling collapse of the
dependence of the average force on velocity for numerical data from 
five different temperatures, as well as for experimental data from 
the work of Gnecco {\it et al.\/}\ \cite{Gnecco_00}.
This collapse 
well confirms the prediction of ramped creep from Eq.\ (1).  
The small deviations from the logarithmic behaviour
at large values of $v/T > 1$ indicate the approach to the 
zero-temperature stick-slip motion \cite{Basel,Helman_94,Hamburg}
The bottom inset of Fig.\ \ref{fig2} shows that, 
in contrast, linear creep does not give a good scaling collapse.

These results can be explained analytically by considering
the thermal activation transition of the tip out a single
metastable well and the distribution of the support's position
$P(R)$ when such a transition occurs \cite{Kurki_72,Garg_95}. 
Translational invariance allows us to concentrate on one such
minimum, of value $\tilde{x}=0$ if $R=0$.
As the support is moved, the barrier 
potential to the next minimum vanishes 
at some critical position $R_c$ \cite{Kurki_72},
determined by 
$\partial \CombinedEnergy/\partial {x_c} = 
\partial^2 \CombinedEnergy/\partial x_c^2 =0$:
\begin{equation}
\frac{2 \pi R_{c}}{a} = \frac{ 2 \pi x_c}{a} 
+ \frac{1}{\Omega_k^2} \sin \left( \frac{2 \pi x_c}{a} \right)
\end{equation}
where $2 \pi x_c /a = \cos^{-1} (-\Omega_k^2 )$.
Note that the support's motion
corresponds to a continuous linear ramping of the potential.
Provided that the velocity of the support is slow enough, there can occur
thermally activated transitions between 
two 
minima before the critical
position $R_c$ is reached.
This is described  by the Kramer's rate \cite{Jung_90}
\begin{equation}
\tau^{-1} = \frac{\Omega^2}{2 \pi \gamma} \;
e^{-\Delta \CombinedEnergy / k_B T} 
\label{Kramer}
\end{equation}
where $\Omega$ and $\Delta \CombinedEnergy$ correspond respectively 
to the instantaneous effective oscillation 
frequency and barrier height. 
Thermally activated transitions
are most likely to occur when the support's position is close
to the critical value. To lowest order in 
the bias $\Bias (t) \equiv  1 - (R(t)/R_{c}) \ll 1$, 
the minimum and maximum
of the potential are easily obtained, and one finds \cite{Kurki_72}: 
\begin{equation}
\Omega = \frac{2\pi}{\sqrt{p_0 p_k}}  \, 
\left( \frac{4 \pi R_c}{a} \right)^{1/4} (1 -\Omega_k^4 )^{1/8} \, 
\Bias^{1/4},
\end{equation}
and,
\begin{equation}
\Delta \CombinedEnergy =  
\frac{2}{3} \, U \, \left( \frac{4 \pi R_c}{a} \right)^{3/2} 
\frac{ \Omega_k^3}{ (1 -\Omega_k^4 )^{1/4} } \, {\Bias}^{3/2}.
\end{equation}
Note the barrier height vanishes as $\Bias^{3/2}$ as anticipated above.
Likewise, the factor $\Bias^{1/4}$ justifies the high-friction limit
($\gamma \gg \Omega$) of the prefactor in Eq. (\ref{Kramer}).

The transition rate represents a measure of the time needed
before a thermally activated transition takes place. In particular, 
$W (R (t)) = \exp  - \int_{t_0}^{t} \tau^{-1} (R(t') dt' $
gives the probability that a transition has not
taken place at time $t$. Due to the exponential character of the 
transition rate, the probability for transition is much greater when
$\Bias \ll 1$ and insensitive 
to the initial support position.  The distribution of the 
support's position at which a transition occurs is then 
conveniently expressed in terms of 
the reduced bias 
$\DimensionlessBias = (\Delta \CombinedEnergy /k_B T)^{2/3}$ as 
\cite{Kurki_72} : 
\begin{equation}
P (\DimensionlessBias) = \frac{3}{2}  \, 
\frac{\DimensionlessBias^{1/2}}{\DimensionlessVelocity} \, 
\exp \left( -\DimensionlessBias^{3/2} - 
(e^{-\DimensionlessBias^{3/2}})/\DimensionlessVelocity\right) 
\label{distribution}
\end{equation}
where the dimensionless velocity 
\begin{equation}
\DimensionlessVelocity (T,v) = 2 \left( \frac{v \, \gamma \, p_0^2
     U}{k_B T a}
\right) 
\frac{\Omega_k^2} { ( 1- \Omega_k^4)^{1/2}}
\label{x_t_v}
\end{equation}
is essentially 
a function of the temperature, microscopic
friction coefficient and velocity of the support. The average of
the distribution $\langle \DimensionlessBias \rangle$ is then found from 
Eq.\ (\ref{distribution}), with value 
$\langle \DimensionlessBias \rangle = 
|\ln \DimensionlessVelocity|^{2/3} + 
{\cal O} (1/\ln \DimensionlessVelocity)$ in the
limit $\DimensionlessVelocity \ll 1$ 
\cite{Kurki_72,Garg_95}.

Once 
the average support's position at which a slip event
occurs is known, it is straightforward to calculate the 
average lateral force as the integral of the instantaneous
force over a cycle of the stick-slip motion \cite{note3}; 
the form of Eq.\ (1) then follows:
\begin{equation}
F = F_c - \Delta F | \, \ln \DimensionlessVelocity |^{2/3},
\label{r_creep}
\end{equation}
where the constants \cite{HERTZ_MODEL}
\begin{equation}
F_c = \tilde{\SpringConstant} \left( R_c - \frac{a}{2} \right),
\label{fc}
\end{equation}
and,
\begin{equation}
\Delta F = \frac{\pi U}{a} \left( \frac{3}{2} 
\frac{k_B T}{U} \right)^{2/3}
\left( \frac{ (1- \Omega_k^4 )^{1/6} }{1+\Omega_k^2} \right). 
\label{deltaf}
\end{equation}
With the parameters used in Fig.\ 2, Eq.\ (\ref{x_t_v}), Eq.\ (\ref{fc})
and Eq. (\ref{deltaf}) predict values of $F_c = 0.55 \, \mbox{nN}$, 
$\Delta F/T^{2/3} = 1.91 \times 10^{-3} \, \mbox{nN/K}^{2/3}$ and 
$T \DimensionlessVelocity /v = 3.9 \times 10^{-2} \, \mbox{K-sec/nm}$. 
This compares well to the respective values of $(0.54 \pm 0.01) \, 
\mbox{nN}$, $(1.82 \pm 0.02) \times 10^{-3} \mbox{nN/K}^{2/3}$
and $(4.5 \pm 0.2) \times 10^{-2} \mbox{K-sec/nm}$ 
extracted from the numerical and experimental 
work, confirming that the ramped creep regime well describes the motion of
the tip.  

 From the distribution function, other properties can be calculated.  
In particular, a directly accessible
experimental quantity is the fluctuation of the maximal force $F_m$, 
the force just
before a slip event occurs.  
Rewriting the 
bias as 
$\Bias = 1 -
F_m/(\tilde{\SpringConstant}R_c)$ allows us to
calculate the distribution of the force 
maxima.  The analytic form is shown in Fig.\ 3, together with numerical
data.  The agreement between the numerical results and theoretical
prediction is very good.  
Finally, we note that it is also straightforward to redo the
above calculation for the restricted regime in which
linear creep takes place.  
This regime is limited to very small velocities, 
less than $0.1 \mbox{nm/sec}$:
an
expansion of $\CombinedEnergy(R,x)$ around 
$\tilde{x}=0$ 
and $\tilde{x}=a$ 
yields the condition
$v < (4\pi a/\gamma p_0^2)
(k_B T/k) \exp (-f_c U/ k_B T)$, where $f_c > 0.5$ is the value of
the bias at which a linear expansion becomes inappropriate. 

To test our results experimentally, it
would be particularly
valuable to consider an extended range of temperatures and velocities,
obtaining not only the dependence of the friction force on those
quantities, but the distribution function of such forces as well.  
The parameters extracted from such comparisons would provide direct,
feasible, and simple access to the fundamental description of friction
at the atomic scale.

We thank Peter Gr\"utter for useful discussions.  This work was supported
by the Natural Sciences and Engineering Research Council of Canada, and
{\it le Fonds pour la Formation de Chercheurs et l'Aide \`a la Recherche
du Qu\'ebec\/}.

\begin{figure} 
\caption{
(a) Schematic representation of the combined tip-surface 
potential just after a slip event. At this point, the energy
barrier is high and thermally activated transitions improbable.
(b) As the support is moved, the energy barrier diminishes and 
transitions become more likely. Inset (c) shows 
typical stick-slip behaviour
of the instantaneous friction force as the support is moved at
velocity $v = 25 \, \mbox{nm/sec}$ for a temperature 
$T=293 \, \mbox{K}$. Inset (d) shows the average friction force
for different temperatures. 
Open circles, squares, diamonds,
triangle-ups and triangle-downs correspond respectively to temperatures
$T = 53, 133, 213, 293$ and $373 \, \mbox{K}$, closed circles are
the experimental data of Ref.\ \protect \cite{Gnecco_00}. The 
units 
of velocity v 
are 
nm/sec. }
\label{fig1}
\end{figure}

\begin{figure}
\caption{
Justification of the scaling form, Eq.\ (\ref{2/3}). 
The upper inset shows the value
$F_c$ as extracted from the data of the friction force shown in 
Fig.\ \ref{fig1} for different temperatures between $53 \mbox{K}$ and
$373 \mbox{K}$ and fixed ratio $T/v = 1 \, \mbox{K/(nm/sec)}$. 
The lower inset shows
significantly worse scaling for linear creep,
using 
$\ln (v/T)$.  The units of velocity v are nm/sec temperature is in
Kelvin.
} 
\label{fig2} 
\end{figure}

\begin{figure} 
\caption{
Normalised distribution, with statistical error bar, of the maxima of the 
friction force $F_m$ at temperature $T=293 \, \mbox{K}$ and velocity
$v = 25 \, \mbox{nm/sec}$. The result from the simulation is well 
reproduced by the theoretical distribution, Eq.\ (\ref{distribution}).
} 
\label{fig3} 
\end{figure}

\end{multicols}


\newpage
\begin{figure}[tbh]
\epsfxsize=6in \epsfysize=6in
\epsfbox{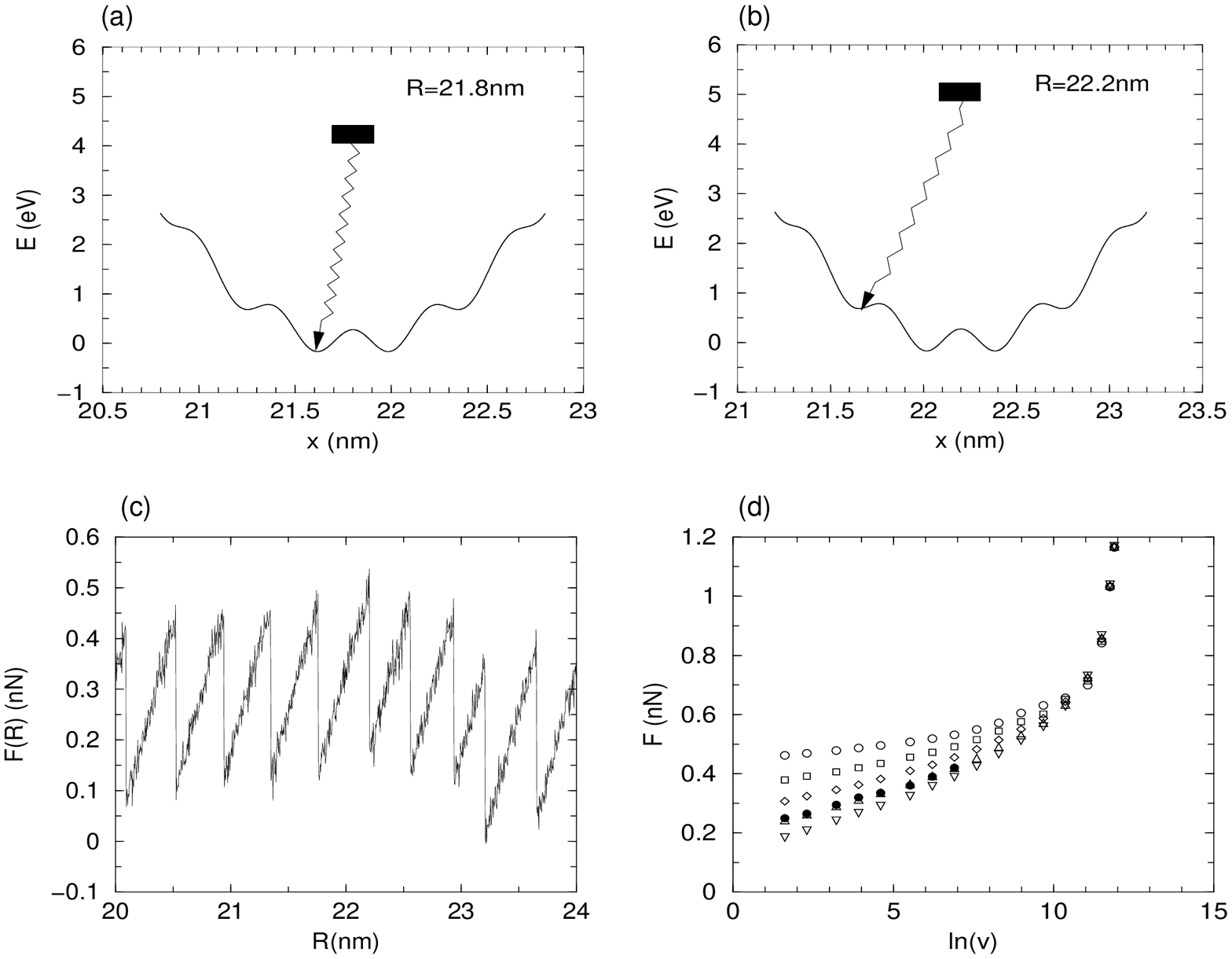}
\end{figure}

\begin{figure}[tbh]
\epsfxsize=6in \epsfysize=6in
\epsfbox{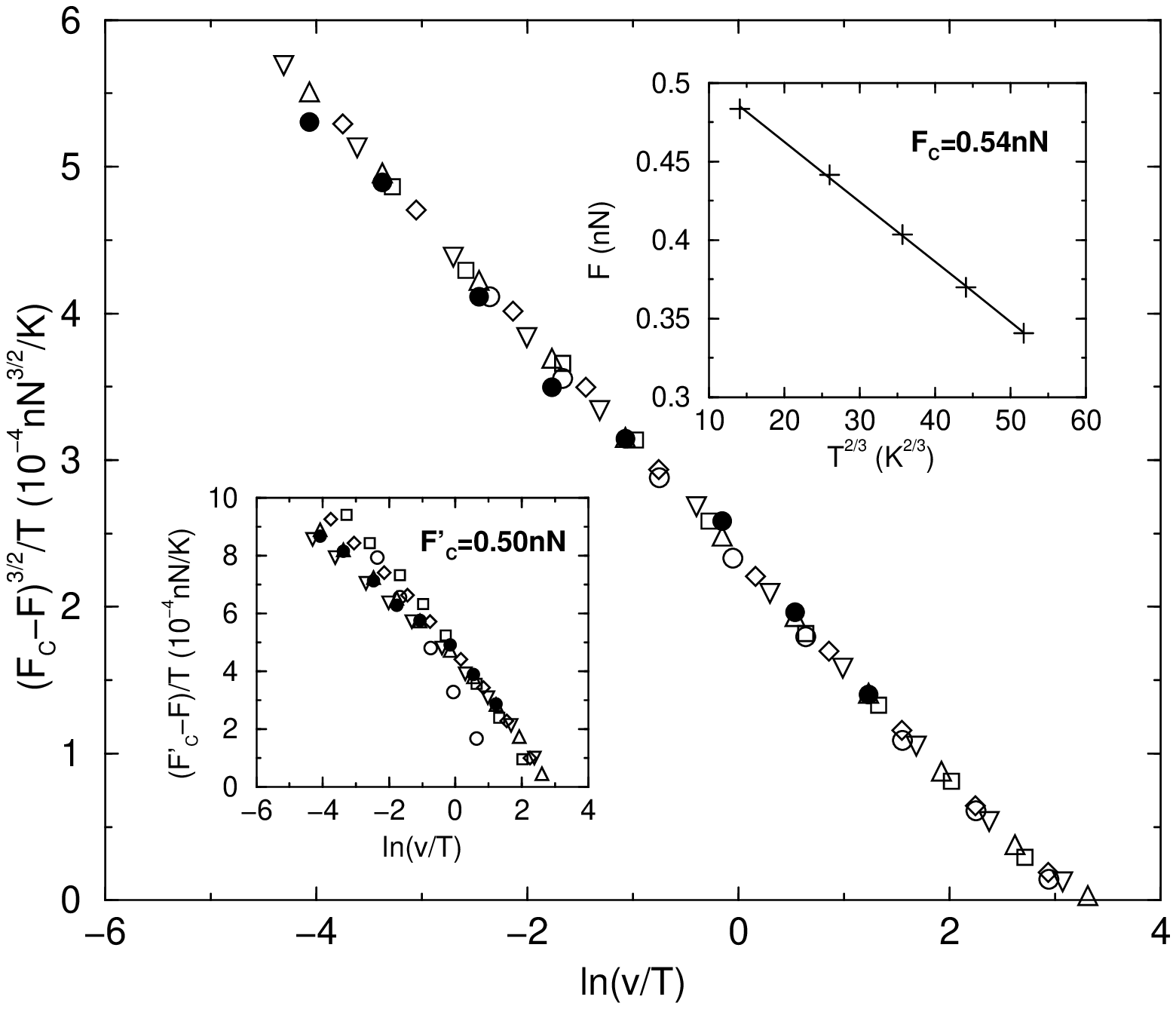}
\end{figure}

\begin{figure}[tbh]
\epsfxsize=6in \epsfysize=6in
\epsfbox{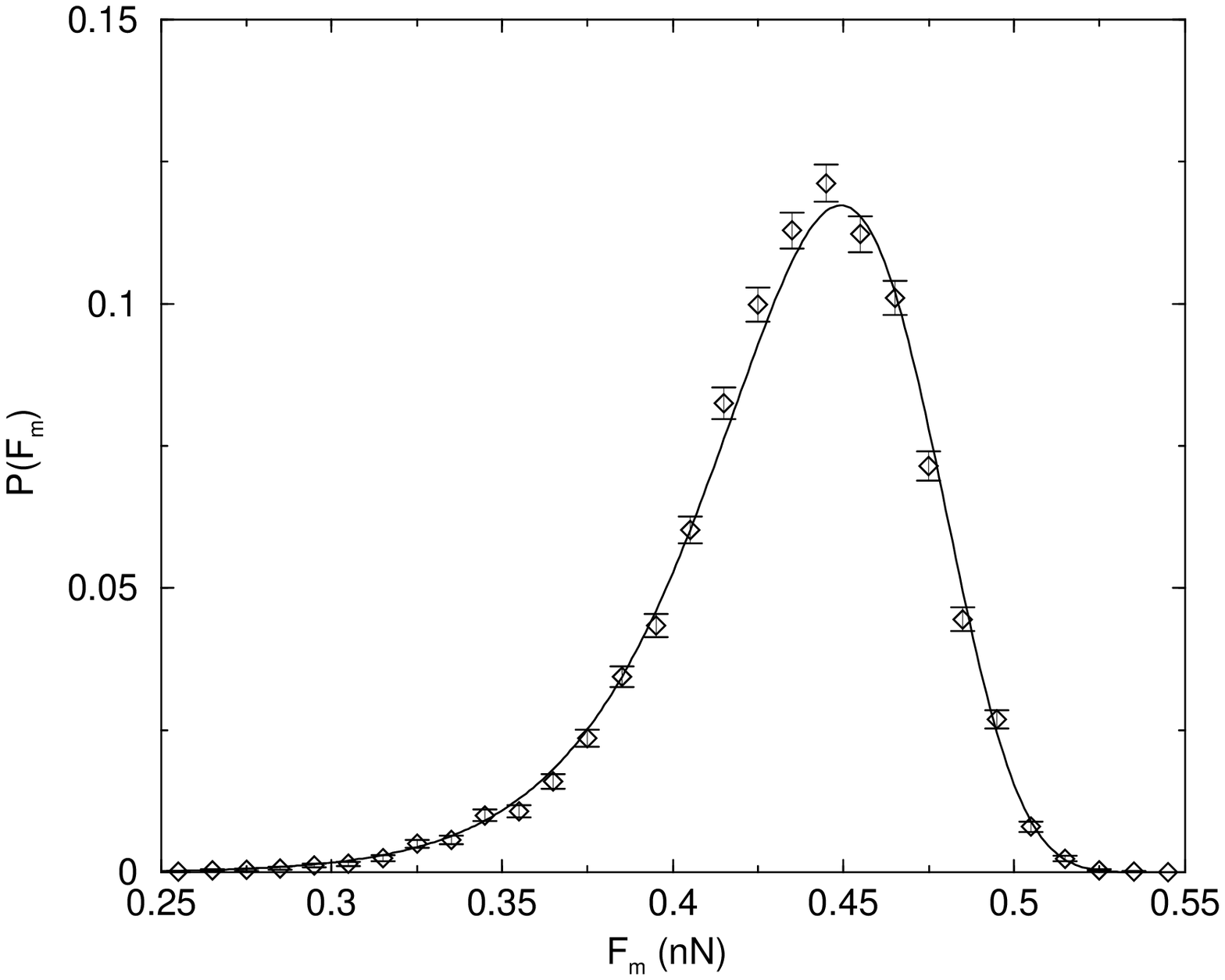}
\end{figure}


\begin{thebibliography}{99}
\bibitem{GAMMA} Of course, $\gamma$ is directly related to the
coefficient of sliding friction. Calculations of the friction coefficient
for adatoms may be found in e.g., Ref.\ \cite{Persson_98}.
The ramped creep regime 
is valid up to
velocities of order 
${\cal O} (F_c/M\gamma)$, where $F_c$ is the
constant force in Eq.\ (\ref{2/3}). 
At this point, 
the tip hardly feels the barrier potential and hence moves without stick
and slip, 
and the
Kramer's activation approach is not valid \cite{Hamburg}.
\bibitem{NEW_REFERENCE}
Some parameters, such as $\gamma$, can have a weak temperature dependence.
Typically, this is negligible, corresponding to a variation of less than
10\% over hundreds of degrees of temperature change.  In any event, the
scaling forms we propose involve, for example, that temperature-dependent
$\gamma$.
\bibitem{note3} 
If the elastic constant between tip and support is small enough that
$\langle R \rangle \geq 2 \, a$,  transitions of the tip over more
than one lattice spacing can take place. The generic behavior
described by Eq.\ (\ref{2/3}) is not modified, except in the academic
case of extremely small $\Omega_k$.
\bibitem{HERTZ_MODEL}
The Hertz model \cite{Persson_98}, which neglects activated processes,
relates the frictional force to $[C(T)F_L]^{2/3}$, where $C$ is inversely
proportional to an elastic modulus, and $F_L$ is the applied load.
This temperature dependence is negligible compared to that obtained herein
from the consideration of activated processes, since the temperature
dependence of $C$ is small \cite{NEW_REFERENCE}.
\end{thebibliography}
\end{document}